\documentstyle[12pt,twoside,fleqn,espcrc1,epsf,psfig]{article}

\newcommand{\be}{\begin{eqnarray}}
\newcommand{\ee}{\end{eqnarray}}
\newcommand{\tr}{\rm tr}

\newcommand{\dirac}{\not\!\partial}

\newcommand{\AmS}{{\protect\the\textfont2
  A\kern-.1667em\lower.5ex\hbox{M}\kern-.125emS}}
\title{ Mean Field, Instantons and Finite Baryon Density}

\author{Momchil Velkovsky  {
Brookhaven National laboratory,\\
P.O.Box 5000, Upton, NY 11973-500, USA\\
momchil@bnl.gov}}

\begin{document}
\maketitle

\begin{abstract}
Instantons create a non-local interaction between the quarks, which at
finite baryon density leads to the formation of a scalar diquark
condensate and color superconductivity.  A mean field approach leads
to a self-consistent description of the $\langle\bar q q \rangle$ and
$\langle q q \rangle$ condensates and shows the inevitability of a BCS
type instability at the Fermi surface. The role of the rearrangement
of the instanton ensemble for the QCD phase transitions is also
discussed.

\end{abstract}

\section{Introduction}

In the last two decades, instantons were shown to give an explanation
of wide range of  non-perturbative phenomena - from the binding of the 
lightest
hadrons and the formation of the chiral condensate to the restoration
of the chiral symmetry at temperatures of about 150 MeV and the
formation of quark gluon-plasma \cite{SS_97}. Although the detailed 
understanding of the QCD vacuum and of phenomena like the color 
confinement still remains elusive, lattice studies tend to show that
the instantons are the principal non-perturbative configurations in
QCD and their role the phenomena mentioned above can be directly
observed on the lattice \cite{latt}. Unfortunately, one area where the
lattice studies are very difficult due to the complex action is the
finite baryon density QCD. It seems that the instanton-inspired models
are some of the very few methods that work in this  area, and
recent studies \cite{WRA,BR,we} have shown an exciting range of new
QCD phases and phenomena like the color superconductivity. One can try
to follow faithfully the instanton  approach starting directly
from the QCD Lagrangian instead of just considering another 4-fermion
model interaction. Here we use a mean field treatment of the effective
interactions for the case of three colors $N_c=3$ and two flavors
$N_f=2$. We  derive the grand canonical potential $\Omega$ at finite
baryon chemical potential $\mu$, containing both the chiral
$\langle\bar q q \rangle$ and the color diquark $\langle q q \rangle$ 
condensates. We use the exact finite  $\mu$ instanton formfactors
and show that the gap equations reveal a BCS like instability at the
Fermi surface, which leads to a diquark condensation. On the other hand
there is a critical coupling for the existence or the chiral
condensate, which  leads to the restoration of chiral symmetry at high 
densities. 

\section{Effective interaction and the finite $\mu $ formfactors}

The method of deriving an effective $2N_f$-fermion interaction
starting from the QCD action was
first developed in \cite{DP} by D. Diakonov and V. Petrov. It consists of
several steps that are sketched below. All momenta and coordinates are 
Euclidean and we use the Euclidean $\gamma$ matrices.
Starting from the QCD partition function,
\be
 Z=\int{\cal D}\psi{\cal D} \psi ^\dagger {\cal D}A \exp({S_{QCD}})=\int{\cal
  D}A  \det(\not\!\!D) \exp({S_{gauge}}),
\ee
integrate out the fermions to obtain the fermion determinant and
separate the low-eigenvalue part from the high-eigenvalue part, by
introducing an arbitrary scale 
\be
 \det(\not\!\!D)= {\rm Det^{low}} {\rm Det^{high}}.
\ee
Consider the classical gauge configuration to consist of an
instanton--anti-instanton ensemble. When sufficiently diluted it can
be approximated by the sum ansatz
\be
A=\sum_{k \in I,\bar I} A_k + \delta A.
\ee
 The only collective effects are produced by the zero mode part of the
fermion determinant. 
It consists of 
\be
T_{I\bar I}=\int dt d^3{\bf x} 
  \phi_I^\dagger(x-z_I)\dirac \phi_{\bar 
I}(x-z_{\bar I}),
\ee
where $\phi$ are the fermion zero modes. They are known also in the
case of finite baryon chemical potential \cite{Abr}
\be
\psi_0=-i\frac {\rho}{
2\pi \sqrt {\rho^2+{x}^2}}{e^{\mu\,t}}\sigma_{{\nu}}\left (
\partial_{{\nu}}-{\frac {\partial_{{\nu}}(1+{(\rho/x)}^{2})}
{1+{(\rho/x)}^{2}}}\right ){e^{-\mu\,t}}
{1 \over x^2}\left(\cos(\mu\,r)+{\frac {t}{r}\sin(\mu\,r)}\right
)\epsilon_R,
\ee
where $\rho$ is the instanton radius.
 
\begin{figure}
\begin{center}
\leavevmode
\hbox{\psfig{file=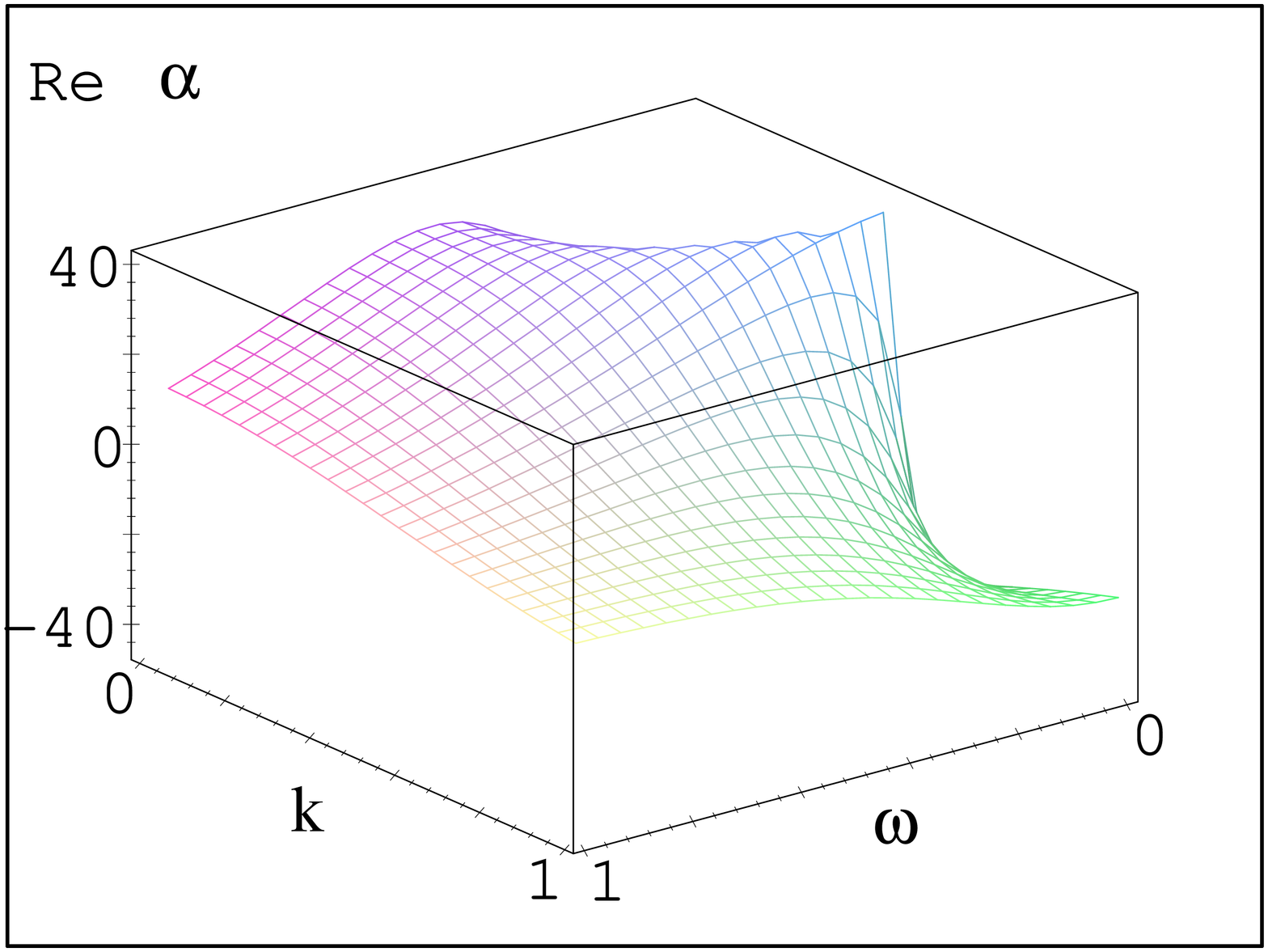,angle=-90,width=7 cm}}
\hbox{ \hskip 0.5 cm \psfig{file=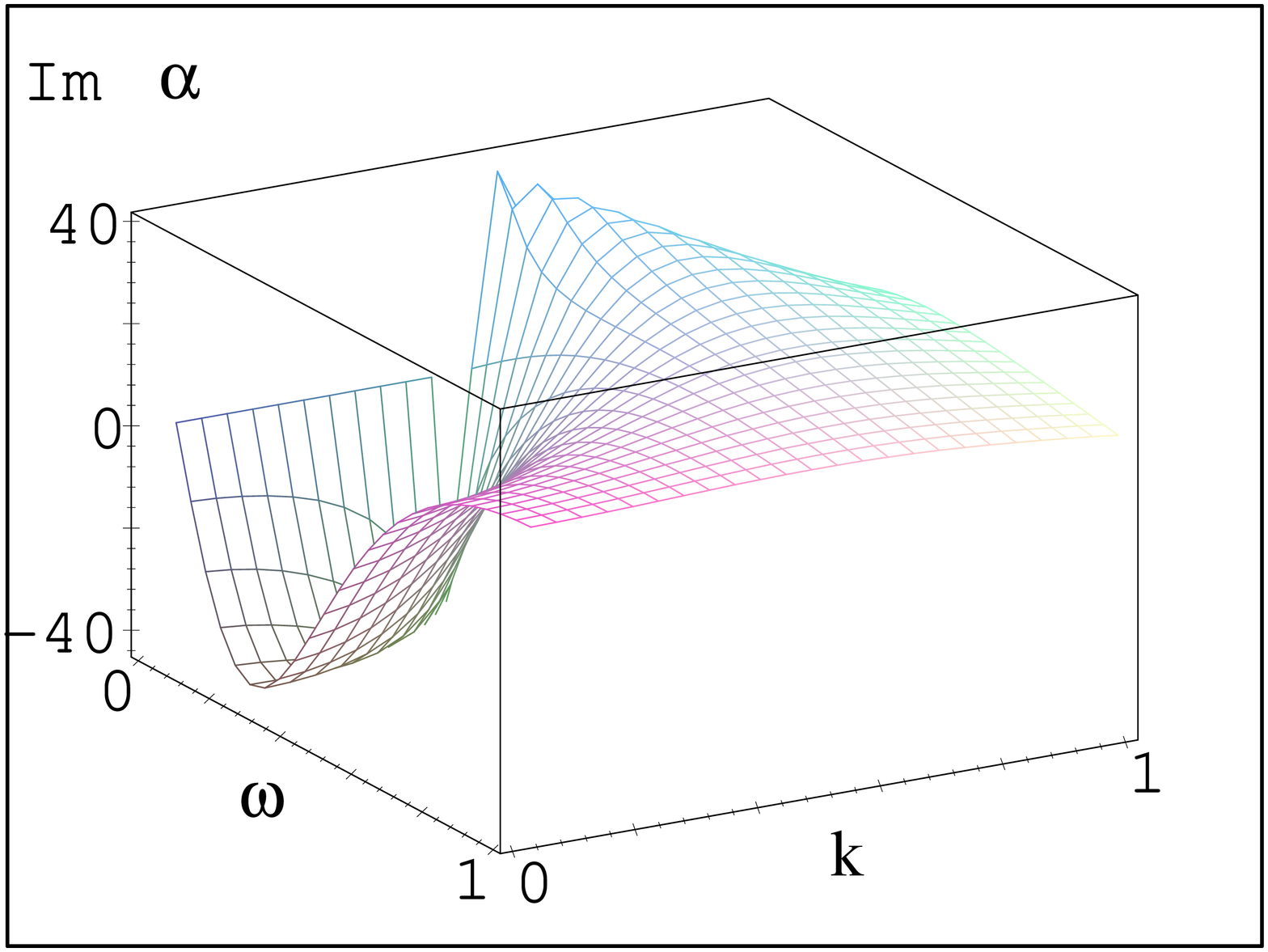,angle=-90,width=7 cm} }
\end{center}
\vskip -0.5 cm
\caption{
The real and imaginary parts of the formfactor at
$\mu=0.5/\rho$. $\omega$ and $k$ are also in units of $1/\rho$.
}
\label{ff}
\end{figure}

Next, reintroduce the fermions, but this time as free fields.
Reproduce the zero mode determinant by calculating a Green's function 
with $N_f (N_++N_-+N_m)$ external legs and using the Wick's
theorem to  do the contractions.
\be
Z=\int d\psi d \psi ^\dagger {\exp \{\int d^4 x 
 \psi ^\dagger i \not\!\!\partial  \psi\} \over 
N_+!N_-}\prod_{I=1}^{N_+} \theta_+
 \prod_{\bar I=1}^{N_-} \theta_- ,
\ee
where in the two flavor case
\be
\theta_+=\int d z_I d\Omega_I \prod_{f=1}^{2}\big[\int
d^4x \psi^
\dagger_f(x) i\not\!\!\partial \phi_I (x-z_I)
\int d^4y\phi_I^\dagger(y-z_I)i\not\!\!\partial 
 \psi_f (y)\big],
\ee
and the integrals are over the collective coordinates of the instantons.
To exponentiate the fermion terms, apply inverse Laplace
Transformation then go to momentum  space and do the integrations over
the center coordinates 
and the orientation angles of  the instantons and anti-instantons.
The formfactor $\alpha(\omega,k;\mu)$ in the resulting non-local
effective Lagrangian is expressed through the Fourier transform of
the fermion zero modes. The exact formula depends on the modified
Bessel Functions ${\rm I}_0(\rho\omega^\pm/2)$ and  
${\rm K}_0(\rho\omega^\pm/2)$ and their derivatives, where 
$\omega^\pm=\sqrt{\omega^+(k\pm\mu)^2}$. The rather
lengthy expression shows that the formfactor is centered at the Fermi 
surface and has a singularity there, but remains finite. Its real 
and imaginary parts are shown on Fig. \ref{ff}. 

For the mean field treatment below, it is convenient to write the
effective interaction known as 't Hooft Lagrangian \cite{'tH}
as a {\em sum} of the S, T and U channels of the
${\cal T}$ matrix in a tree approximation by doing the appropriate Fierz 
transforms. The meson channels (S+T)
have color singlet and color octet terms, the scalar ones are
attractive, while the pseudoscalars are repulsive:
\be 
\label{l_mes}
{\cal L}_{mes} &=& G\left\{ {1\over 8 N_c^2}
 \left[(\bar\psi \tau^- \psi)^2+
 (\bar\psi \tau^- \gamma_5 \psi)^2 \right]+
 {N_c-2\over 16 N_c^2 (N_c^2-1)}
 \left[(\bar\psi \tau^- \lambda^a \psi)^2 \right. \right. \nonumber \\
 & &  \left. \left. \mbox{}
   \!   \!
 +(\bar\psi \tau^- \lambda^a \gamma_5 \psi)^2 \right]  
-{1\over 32 N_c (N_c^2-1)}(\bar\psi \tau^-  \sigma_{\mu \nu}
\lambda^a \psi)^2 \right\}.
\ee
The diquark channels contains both color antisymmetric ${\bf \bar 3}$
and  symmetric ${\bf 6}$ terms, the scalar and the tensor are attractive:

\be 
\label{l_diq}
{\cal L}_{diq} &=& 
G \left\{
 -{1\over 8 N_c^2 (N_c-1)}
 \left[ (\psi^T C \tau_2 \lambda_A^a \psi)
        (\bar\psi\tau_2 \lambda_A^a C \bar\psi^T)  
       +(\psi^T C \tau_2 \lambda_A^a \gamma_5 \psi)
        (\bar\psi \tau_2 \lambda_A^a \gamma_5 C \bar\psi^T) \right]
\right.\nonumber\\
 & &  \left. \mbox{}
   \!   \! 
        +{1\over 16 N_c^2 (N_c+1)}
        (\psi^T C \tau_2 \lambda_S^a \sigma_{\mu \nu} \psi)
        (\bar\psi \tau_2 \lambda_S^a \sigma_{\mu \nu} C \bar\psi^T) 
        \right\},
\ee 
where $C$ is the charge conjugation matrix, $\tau_2$ is the anti-symmetric 
Pauli matrix, $\lambda_{A,S}$ are the anti-symmetric (color ${\bf \bar
3}$) and symmetric (color {\bf 6}) color generators (normalized in an 
unconventional way,
${\rm}tr(\lambda^a\lambda^b)=N_c\delta^{ab}$, in order to facilitate the 
comparison between mesons and diquarks). When written in momentum 
representation, each fermion has a formfactor $\sqrt{\alpha}$ attached
to it. The overall coupling constant G comes from the exponentiation
of the fermion terms through the inverse Laplace transformation. One
can do a saddle point approximation of this integral by varying G and
thus relate it to the parameters of the instanton ensemble - total
density $n_{inst}$ and average radius $\rho$.

\section{Mean field approach}

\begin{figure}
\begin{center}
\leavevmode
\centerline{\psfig{file=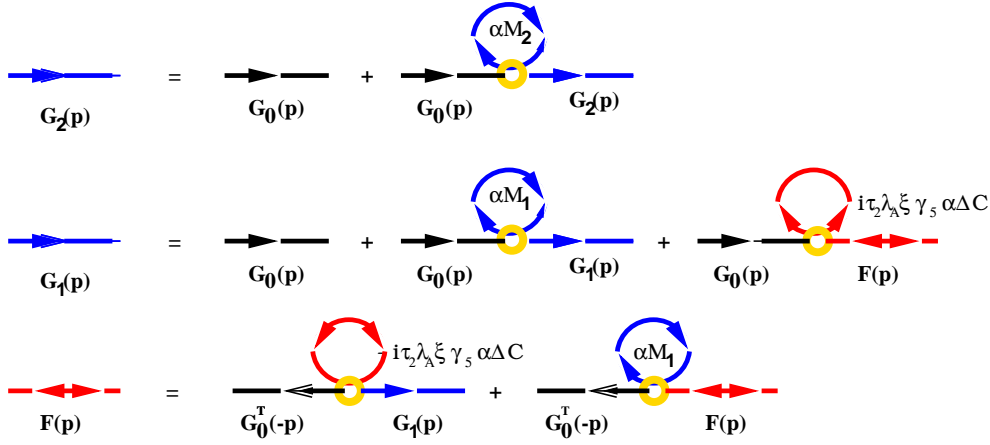,width=13. cm}}
\end{center}
\vskip -0.5 cm
\caption{Hartree-Fock equations for the propagators. The effective
masses $\tilde M_1,\tilde M_2$ and the gap $\tilde \Delta$ are
proportional to the
corresponding chiral and color condensates.}
\label{hf}
\end{figure}

There is a variety of mean field methods, that yield equivalent
results. One of the most convenient is the use of a bilocal effective 
action $\Gamma[G,F,\bar F]$ \cite{JT,LP,DL}, which depends on the propagators
$G(x,y)=\langle T \psi(x) \bar \psi(y)\rangle,
F(x,y)=\langle T \psi(x) \psi^T(y)\rangle,$ and
$\bar F(y,x)=\langle T \bar\psi^T(x) \bar \psi(y)\rangle.$
In momentum representation one obtains:
\be
\Gamma[G,F,\bar F]=-1/2 \tr \ln[ G(-p)]-1/2 \tr \ln[G(p)+
F(p)(G^T(-p))^{-1}\bar F(-p)]\nonumber \\ +
1/2\tr[G_0^{-1}(p)G(p)+G_0^{-1}(-p)G(-p)-2]+
V[G,F,\bar F].
\label{Gamma}
\ee
Assuming the breakdown  both of the chiral and the color symmetry
through scalar color singlet $ \langle \bar \psi \psi \rangle$ and 
color ${\bf \bar 3}$ $ \langle \psi^T C \gamma_5 \tau_2 {\vec \xi} \cdot
{\vec\lambda_A }\psi \rangle $ condensates, one can write the relevant
for the mean field approach terms of the potential
$V$ in the zeroth meson-diquark loop approximation 
\be
V[G,F,\bar F]=  G {1\over 8 N_c^2}
(\tr[G_1+G_2]|\alpha|)^2+G{N_c-2\over 16 N_c^2
(N_c^2-1)}(\tr [\lambda_8(G_1+G_2)|\alpha|])^2\nonumber \\
+G{1\over 8 N_c^2 (N_c-1)}
\tr [ C \gamma_5 \tau_2 {\vec\lambda_A} F\alpha(p)]\tr 
[ C \gamma_5 \tau_2 {\vec\lambda_A}\bar F\alpha(-p)].
\ee
The unit vector $\vec \xi$ determines the direction of the diquark
condensate in the space of $\lambda_A$. Without restrictions we can
chose it to point towards $\lambda_2$. The breakdown of the color
symmetry leads to the formation of two different chiral condensates, 
one involving the quarks of the first two colors, and the second -- for
the quarks of the third color. Correspondingly, we split the
propagator $G=G_1+G_2$, with $G_1$ containing the projector $P$ to the
upper left $2\times2$ square of the SU(3) matrices, while $G_2$ containing
the projector to the lower right entry. Because both $G_1$ and $G_2$
mix with the $\lambda_8$ matrix, one has to include the color octet
interaction term, which contains $\lambda_8$. (Please, note again the
non-standard normalization of the $\lambda$ matrices.) Varying
$\Gamma$ one obtains the Hartree-Fock equations for the propagators.
In a picture representation they are shown in Fig. \ref{hf}.  

The solution of these equations is
\be
G_2&=&{-(i\not p^-+m_2)(1- P) \over (p^-)^2 +m^2},
G_1={(i\not p^++m_1) P \over (i \not p^- -m_1)(i \not p^+
+m_1)+\Delta^2},\nonumber\\
F&=&{-i\tau_2\lambda_2C\gamma_5\tilde\Delta\alpha(-p) 
\over (i \not p^+ +m_1)(i \not p^-
-m_1)+ \Delta^2},
\ee
where $\not p^\pm=\gamma_0(\omega\pm i\mu)+ \vec \gamma \vec k $ and
$m_{1,2}= m_0+M_{1,2}$, with $m_0$ being the current quark mass and
$M_{1,2}= \alpha(\omega,k,mu)\tilde M_{1,2},\Delta_{1,2}=
\alpha(\omega,k,mu)\tilde\Delta{1,2}$.
Substituting into (\ref{Gamma}) one obtains the grand canonical
potential
\be
\Omega(\tilde M_1,\tilde M_2,\tilde \Delta;\mu) =4\int{d^4p\over (2\pi)^4}
 \ln[((\omega_1^+)^2+\Delta^2)((\omega_1^-)^2+\Delta^2)]+
2\int{d^4p\over (2\pi)^4}
 \ln[(\omega_2^+)^2(\omega_2^-)^2] \nonumber \\
-16\int{d^4p\over (2\pi)^4}{m_1M_1(s_1+\Delta^2-\mu^2)+
\Delta^2(s_1+\Delta^2+\mu^2)\over
((\omega_1^+)^2+\Delta^2)((\omega_1^-)^2+
\Delta^2)}-8\int{d^4p\over (2\pi)^4}{m_2M_2(s_2-\mu^2)
\over (\omega_2^+)^2(\omega_2^-)^2}\nonumber \\
+{8\over 9}G\left[\int{d^4p|\alpha|\over
(2\pi)^4}\left({2m_1(s_1+\Delta^2-
\mu^2)\over ((\omega_1^+)^2+\Delta^2)((\omega_1^-)^2+\Delta^2)}+{m_2(s_2-\mu^2)
\over (\omega_2^+)^2(\omega_2^-)^2}\right)\right]^2\nonumber \\
+{1\over 9}G\left[\int{d^4p|\alpha|\over
(2\pi)^4}\left({m_1(s_1+\Delta^2-
\mu^2)\over ((\omega_1^+)^2+\Delta^2)((\omega_1^-)^2+\Delta^2)}-{m_2(s_2-\mu^2)
\over (\omega_2^+)^2(\omega_2^-)^2}\right)\right]^2\nonumber \\
+{2\over 3}G\left[\int{d^4p|\alpha|\over (2\pi)^4}{\Delta(s_1+\Delta^2+\mu^2)
\over ((\omega_1^+)^2+\Delta^2)((\omega_1^-)^2+\Delta^2)}\right]^2,
\ee
where $(\omega_{1,2}^\pm)^2=\omega^2+(\sqrt{k^2+m_{1,2}^2}\pm\mu)^2$,
$s_{1,2}=\omega^2+k^2+m_{1,2}^2$. By differentiating with respect to
$ \tilde M_1,\tilde M_2$ and $\tilde \Delta$ one can obtain a system
of coupled gap equations, which can be solved numerically for each
value of $\mu$. An easier approach for studying the phase diagram of
the system is to plot $\Omega$ as a function of two of the parameters
for different values of $\mu$ and a fixed third parameter. The minima
will correspond to the different possible phases and the phase
transitions correspond to two minima having an equal value (first
order), or joining together (second order). But even without a computer,
one can study the limiting cases -- pure chiral condensate ($ \tilde
M_1=\tilde M_2,\tilde \Delta=0 $), or pure diquark condensate  ($ \tilde
M_1=\tilde M_2=0$). The corresponding gap equations have the following
form:
\be
1&=&{1\over 12}G\int{d^4p\over (2\pi)^4}{(s_1+\Delta^2+\mu^2)|\alpha|^2
\over ((\omega_1^+)^2+\Delta^2)((\omega_1^-)^2+\Delta^2)},\\
1&=&{2\over 3}G\int{d^4p\over (2\pi)^4}{(s_1-\mu^2)|\alpha|^2
\over(s_1-\mu^2)^2+4\mu^2\omega^2 },
\ee
The formfactor $|\alpha|$, which  enters
in $M_1, M_2, \Delta$, although having a cusp at $k=\mu$ is also
continuous, so that for the qualitative picture one can substitute it
with a sharp cutoff around the Fermi surface. If we remove the gap
$\tilde\Delta$ the first equation develops a logarithmic divergence of
the BCS type and can not be satisfied for any G:
\be
1={1\over 96 \pi^2}GF_1\left[\int_{\mu-\epsilon}^{\mu+\epsilon}{k^2 dk\over
k+\mu}+\int_{\mu}^{\mu+\epsilon}{k^2 dk\over
k-\mu}+\int_{\mu-\epsilon}^{\mu}{k^2 dk\over \mu-k}\right].
\ee
The last two terms, which are divergent correspond to quarks and holes
close to the Fermi surface $k=\mu$. The second equation, however, is of
Nambu type
\be
1={1\over 6 \pi^2}GF_2 \int_{(\mu^2-m^2)^{1/2}}^{mu+\epsilon}{k^2 dk
\over \sqrt{k^2+m^2}}
\ee
and needs a minimal critical coupling for a solution to exist even at
finite $\mu$. Note
also that the Fermi momentum $k_F=\sqrt{\mu^2-m^2}$.
Similar conclusions were reached in \cite{WRA} using a different
method and ad-hoc formfactors.

\section{Phase Transitions}

Although in the picture drawn so far there can be phase transitions,
they may be very dependent on the particular scenario, e.g. in
\cite{WRA} the chiral restoration transition is due mainly to the
formfactor which is centered not at $k=\mu$, but at $k=0$, so that at
sufficiently large chemical potential, it cuts off effectively the
interaction. Another possibility is the Debye screening of the
instantons in the plasma phase, which also leads to decrease in the
effective  four-fermion coupling. Studies of the {\em finite temperature}
chiral restoration transition \cite{SS_96,VS_97}, has shown the importance
of the  correlations in the instanton ensemble, which grow with
$T$. One can take into account these correlations in the mean field
approach by considering the tightly bound and polarized instanton --
anti-instanton pairs as new objects called {\em molecules}, which form
a new component. The phase transition is characterized by complete
pairing of all instantons into molecules.
The role of the molecules is to deplete the random component, which is
responsible for the collectivization of the zero modes and it also
provides additional four-fermion interactions.

At finite {\em chemical potential}, two  mechanisms for suppressing 
$\langle \bar\psi \psi\rangle$ may exist: the  depletion of the 
random component by the molecules, and the competition from 
$\langle\psi \psi\rangle$ - i.e. instantons previously 
engaged in the infinite ``vacuum quark clusters''
(the collectivized zero modes)  become engaged in the 
``matter diquark chains''.

If $\mu>>m_s$, we should consider the $s$ quark as a light one and study the 
phase diagram as a function of its mass too. As recent papers has
shown \cite{BR}the phase structure in the $N_f=3$ case is richer and more 
exotic. 
There is the possibility of many new condensates, and patterns of
symmetry breaking, i.e. ${SU(3)_c}\times{SU(3)_f}\to SU(3)_d$.
The color superconductivity is complete, because there is a gap 
for all quarks.
In this case the instantons can not form alone a diquark 
condensate, but the molecules can. Their contribution to the
diquark binding may be not a correction,
but the main one.

\section{Conclusion}

The mean field approach to the instanton induced quark interaction has
proved itself as a reliable method to study the richness of QCD
phenomena at finite baryon density, a situation in which very few
other methods seem to work. It has shown that the expectations of a
BCS type of superconductivity transition were correct, and
simultaneously they show how the chiral symmetry is restored. However,
if one wants to make this approach more realistic, he should take into
account the correlations and the restructuring of the instanton
ensemble, something that has been shown to play the key role in the
finite temperature phase transition.  

\section{Acknowledgements}

I would like to thank my collaborators Edward Shuryak, Ralf Rapp
and Thomas Sch{\"a}fer for the useful discussions. I also want to
thank D. Diakonov for the healthy criticism and the interesting 
discussions.

\end{document}